\documentstyle[preprint,aps]{revtex}
\begin{document}
\draft
\title{RANDOMNESS, NONLOCALITY AND INFORMATION IN ENTANGLED
 CORRELATIONS}
\author{Krzysztof W\'odkiewicz\cite{aaa}}
\address{Center for Advanced Studies and Department of Physics and
Astronomy, \\
University of New Mexico, Albuquerque, New Mexico 87131, USA \\
and \\
Instytut Fizyki Teoretycznej, Uniwersytet Warszawski
Warszawa 00-681, Ho\.za 69, Poland\cite{aaa}}
\date{\today}
\maketitle
\begin{abstract}
It is shown that  the Einstein, Podolsky and Rosen (EPR)
correlations for arbitrary spin-$s$ and the Greenberger, Horne and
Zeilinger (GHZ) correlations for three particles can be described  by
nonlocal joint and conditional quantum probabilities.
The nonlocality of these  probabilities makes the Bell's  inequalities void.
A description that exhibits the relation between the randomness and
the nonlocality of entangled correlations is introduced.  Entangled EPR and
GHZ correlations are studied using the  Gibbs-Shannon entropy. The nonlocal
character of the EPR correlations is tested using the information
Bell's inequalities. Relations between the randomness, the nonlocality
and the entropic information for the EPR and the GHZ correlations  are
established and discussed.

\end{abstract}


\section{INTRODUCTION}

One of the most fundamental features of quantum correlations is the nonlocal
and the random character of microscopic single events. Nonlocal
randomness differs in a fundamental way from the randomness associated with
the concept  of local realism  which is based on the  assumption
that microscopic physical systems can be described by local objective
properties that are  independent from observation.  Local realism versus the
nonlocal character of quantum mechanics \cite{redhead} has been
described best in the framework of spin-$\frac{1}{2}$  EPR correlations
\cite{epr}. The classical limit of quantum nonlocality has been investigated
in the framework of the generalized EPR entangled correlations involving
arbitrary spin-$s$ \cite{mermin1}. In the limit of large spin these
correlations should exhibit a smooth transition to the classical regime if
the measurements do not resolve the quantum properties of the spin
\cite{garg,mermin2}.

Recently it has been shown by Greenberger, Horne and
Zeilinger (GHZ) \cite{ghz1} that special entangled states involving
three or four particles lead to a much stronger refutation of
local realism.  It has been shown  that  in many-particle correlations only a
{\it single} set of observations is required in order to demolish the
local reality assumption \cite{ghz2}.

It is the purpose of this paper to show that  the EPR spin-$s$
correlations and that the three-particle GHZ correlations can be visualized as
nonlocal correlations between
sequences of completely random numbers $"0"$ and $"1"$ corresponding
to  "no" and "yes" outcomes of the spin polarization analyzers.
The coincidences  between these random numbers are described by   conditional
probabilities which are nonlocal, i.e., are
dependent on the orientations of the spin analyzers. The nonlocality
of these  conditional probabilities makes the Bell's inequality  void,
because
for local realism the existence of a universal, polarization independent
transition probability between the random sequences is essential.

A description that exhibits the relation between the randomness and
the nonlocality of entangled quantum states is well suited to study  the EPR
and the GHZ  correlations using the  Gibbs-Shannon entropy. For  a system
involving  two of three correlated subsystems, the information entropy can
offer a measure
of quantum correlation if the informations of the subsystems
are compared with the total information
of the composed system. The nonlocal character of the EPR and the GHZ
correlations of the subsystems will be tested using the information
Bell's inequalities
involving entropies of the probabilities that can be determined from
the statistics of random "yes" and "no" answers. In such a framework  the
relations between the randomness, the nonlocality
and the entropic information for the EPR and the GHZ correlations will be
established.

This paper is organized in the following way. In Sec. II general properties
of the EPR correlations for arbitrary spin-$s$ are reviewed and the
randomness and the nonlocality of these correlations are discussed in the
framework of the quantum bivalued distribution functions. The randomness and
the information of these EPR correlations is than discussed in the framework
of the information entropy. The information Bell's  inequality for the EPR
correlations involving arbitrary  spin-$s$ is investigated. In Sec. III the
random and the nonlocal properties of the GHZ correlations are discussed. The
randomness and the information of these GHZ correlations are discussed in the
framework of the information entropy. The corresponding information Bell's
inequality for the three particles GHZ correlations is introduced and
investigated.

\section{ENTANGLED EPR CORRELATIONS FOR ARBITRARY SPIN-$s$}

\subsection{ RANDOMNESS AND NONLOCALITY}

The EPR entangled state of the two particles $a$ and $b$, each  with spin $s$
is given by  the following singlet state of the two  spins:

\begin{equation}\label{psi1}
|\psi_{\rm EPR}\rangle = \sum^s_{m=-s} {(-1)^{s+m} \over \sqrt{2s+1}}
|m;-m\rangle.
\end{equation}
where the states $|m_{a};m_{b}\rangle$ correspond to the eigenvalues of the
$z$-components of the two spins. In quantum mechanics, the joint  spin
transmission probability of  detection of such a entangled state by two
linear polarizers, is defined by the following formula:

\begin{equation}\label{qpro1}
p(\vec a; \vec b\;) =\langle \psi_{\rm EPR} | \hat P(\vec a\;) \otimes \hat
P(\vec b\;)|\psi_{\rm EPR}\rangle .
\end{equation}
In this expression  $\hat P(\vec a\;)$ and $ \hat P(\vec b\;)$
are the spin  projection operators of the particles $a$ and $ b$
along the polarization directions $\vec a$ and $ \vec b$.

Following the basic  ideas of a  theory based on local reality, the
transmissions of an arbitrary spin-$s$ through the  linear polarizers $\vec
a$ and $\vec b$ are described by objective realities represented by
normalized  transmission functions $0\leq t(\vec a, \lambda_a) \leq 1$ and
$0\leq t(\vec b, \lambda_b)\leq 1 $, with hidden variables  $\lambda_a$ and
$\lambda_b$.  In a theory based on local hidden variables (LHV),  these
transmission functions are local realities  that  are averaged over the hidden
variables during a detection process.  In the framework of such LHV
theories  the joint spin transmission is given by the following statistical
average of local realities:
\begin{equation}\label{lhv}
p(\vec a; \vec b) = \int d\lambda_a\int d \lambda_b \ P(\lambda_a; \lambda_b)
\ t(\vec a, \lambda_a)
\ t(\vec b, \lambda_b),
\end{equation}
where the hidden parameters are randomly distributed with a  positive
normalized distribution function:

\begin{equation}
\int d\lambda_a \int d \lambda_b P(\lambda_a; \lambda_b)=1, \ \ \  \ \ \
P(\lambda_a; \lambda_b) \geq 0.
\end{equation}
This distribution function is local i.e., it is independent on the polarization
directions $\vec a$ and $\vec b$. For such a local distribution of random
variables, the  LHV transmission probability (\ref{lhv}) evaluated for four
different polarization axes
$\vec a,\ \vec a',\ \vec b,\ \vec b'$ is  restricted by the  Bell's
inequality \cite{bell1}:

\begin{equation}\label{bell1}
-1 \leq  p(\vec a; \vec b) + p(\vec a'; \vec b) - p(\vec a; \vec b')
+ p(\vec a'; \vec b') - p(\vec a\;) - p(\vec b\;) \leq 0 ,
\end{equation}
The original Bell's inequalities have been derived only for spin-$\frac{1}{2}$
, but it has been shown that the inequality in the form (\ref{bell1}) is
valid for correlations involving an arbitrary spin-$s$ \cite{kw1}.

We begin the  discussion of the randomness and the nonlocality in quantum
mechanics using the two-point   distribution that has been introduced and
applied to the discussion of quantum jumps in optical transitions
\cite{kwjhe}.
 Using the spectral decomposition of the spin projector operator: $ \hat P =
\int d \lambda \ \delta(\lambda -\hat P)$
we  rewrite the quantum mechanical joint-probability (\ref{qpro1}) in the
following form:
\begin{equation}\label{clickp1}
p(\vec a; \vec b) = \int d \lambda_a \int d \lambda_b \
P(\vec a \ \lambda_a; \vec b \ \lambda_b)\ \lambda_a \ \lambda_b ,
\end{equation}
where the distribution function is given by the following quantum
mechanical average:
\begin{equation}\label{nonlocp1}
P(\vec a \lambda_a; \vec  b \lambda_b) = \langle \psi_{\rm EPR} |\ \delta
(\lambda_a
 - \hat P( \vec a)) \otimes \delta (\lambda_b - \hat P(\vec b))\ |\psi_{\rm
EPR}\rangle.
\end{equation}
The projection operators $\hat P( \vec a) =  | \vec a \rangle \langle \vec a
|$ and $\hat P( \vec b) =  | \vec b \rangle \langle \vec b|$   corresponding
to the two orientations of the linear polarizers $\vec a $ and $\vec b$ are
obtained by a rotation of the maximum "down" spin states $|m_{a}=-s\rangle$ and
 $|m_{b}=-s\rangle$ by
angles $\tau_a=\frac{1}{2}\theta_a e^{-i\phi_a}$ and
$\tau_b=\frac{1}{2}\theta_b e^{-i\phi_b}$ respectively:
\begin{equation}\label{projection}
| \vec a \rangle  = \exp( \tau_a {\hat S}_{a+} -
\tau^{*}_a {\hat S}_{a-})|-s\rangle \ \ \ {\rm and} \ \ \  | \vec b \rangle
= \exp( \tau_b {\hat S}_{b+} -
\tau^{*}_b {\hat S}_{b-})|-s\rangle ,
\end{equation}
where   ${\hat S}_{a\pm}$ and  ${\hat S}_{b\pm}$ are the spin-$s$ ladder
operators for the particle $a$ and $b$.

   With the help of the distribution function (\ref{nonlocp1}) we have
rewritten the quantum
mechanical joint probability  function (\ref{qpro1}) in a form given by
(\ref{clickp1}) which  has remarkable similarities to
the LHV correlation function given by (\ref{lhv}).  Because the projection
operators can have their eigenvalues equal to 1 or 0, i.e., can represent
only
``yes'' or ``no'' answers, the values $\lambda_a$ and $\lambda_b$ can take
only values
equal to 1 and 0  corresponding to "clicks" at the detectors.
The bivalued distribution given by (\ref{nonlocp1}) is positive
everywhere, but depends on the polarization directions $\vec a$ and $\vec b$.
The
distribution function which depends on the orientation $\vec a$ of the first
analyzer and on the orientation $\vec b$ of the second (possibly even remote)
analyzer is nonlocal. In the framework of   EPR correlations, it is
customary to call an analyzer-dependent distribution function a nonlocal
distribution function.  The
nonlocality of this distribution function makes the Bell's inequality
(\ref{bell1}) void,
because in order to obtain this inequality the existence of a universal,
local
(polarization independent) distribution in the parameters $\lambda_a$ and
$\lambda_b$  (hidden parameters in this case) is essential.

{}From quantum mechanics we obtain that  EPR correlations can be described
by  a  distribution function of the form
given by Eq.~(\ref{nonlocp1}) with the condition that the
statistical distribution of the parameters $\lambda_a$ and $\lambda_b$ is
nonlocal i.e., is  dependent on the polarization direction.

In order to elucidate the statistical and  the random character of the
nonlocal distribution function further, we shall perform a Bayesian
analysis of the joint correlation. The joint distribution can
be written in the following form:

\begin{equation}\label{bayes1}
P(\vec a \lambda_a; \vec  b \lambda_b) =
P(\vec a \lambda_a | \vec  b \lambda_b) P(\vec b\lambda_b).
 \end{equation}
The distribution $P(\vec a \lambda_a | \vec  b \lambda_b)$ is the conditional
of the event $\lambda_a$
(``yes'' or ``no'') to occur under the condition that $\lambda_b$ (``yes'' or
``no'') has occurred and $P(\vec b \lambda_b)$ is the marginal of the joint
distribution. In the following, for notational convenience, we will omit the
polarization directions and denote these functions by $P(i|j)$ and $P(j)$,
where
the indexes $i$ and $j$ will denote $1$ and $0$ for a "yes" or a "no" outcome,
respectively.
{}From the properties of the EPR entangled state (\ref{psi1}) and the
definitions of the projection operators (\ref{projection}) it is easy to
calculate  all of  these distribution functions.
The single-folded distributions are:
\begin{equation}\label{singlemarg}
P(0)= {2s\over 2s+1}
\quad {\rm and} \quad P(1)={1\over 2s+1}.
\end{equation}
Note that they are local i.e., polarization independent.
The nonlocal polarization dependent conditional distribution  is given by
the following matrix:
\begin{equation}
P(\lambda_a| \lambda_b) =
   \left[ \matrix {P(0|0) &P(0|1) \cr P(1|0) &P(1|1)
\cr } \right] ,
\end{equation}
where

\begin{equation}\label{cp1}
P(0|0)  ={1\over 2s}\left( 2s-1 + \left( \sin{\alpha \over 2}
\right)^{4s}\right)
\ \ {\rm  and} \ \  P(0|1) =1-  \left( \sin {\alpha \over 2} \right)^{4s}.
\end{equation}
In these formulas $\alpha$ is the relative angle between the two unit
vectors $\vec a$ and $\vec b$ ($\cos\alpha = {\vec a} \cdot {\vec b}$). From
these joint and conditional probabilities we obtain that the joint spin
transmission function is:
\begin{equation}
p(\vec a; \vec b) = P(1|1)P(1)= \frac{1}{ 2s+1} \left( \sin {\alpha \over 2}
\right)^{4s}.
\end{equation}

{}From the Bayes analysis (\ref{bayes1}) we have the following sum rules
fulfilled by the conditional probabilities:
\begin{equation}
P(0|0)+ P(1|0)=1 \quad {\rm and}\quad  P(0|1)+ P(1|1)=1.
\end{equation}
This result shows that one can regard the
EPR correlations as  correlations of two sequences of random numbers
represented by transmission functions $t(\vec a, \lambda_a) =\lambda_a$  and
$ t(\vec b,\lambda_b)=\lambda_b$ that are jumping between values 0 and 1
(``no'' and ``yes'' answers) for  measurements performed with linear
analyzers.
 These are the only possible outcomes of the transmission experiment. This
positive and nonlocal distribution leads to a simple statistical
interpretation of the spin correlations and of the violation of Bell's
inequality in terms of random numbers 1 and 0 for the transmission
functions.    On each single polarizer, the outcomes
are completely random and the ``yes'' and ``no'' answers occur with
probabilities $P(0)$ for "no" and $P(1)$ for "yes".  The nonlocality of the
EPR correlations shows
up in the fact that these two perfectly random sequences (on the first and
the
second polarizers) are correlated and the correlations are given by
Eq.~(\ref{cp1}).
These formulas predict that the EPR entanglement can be understood as a
nonlocal correlation between the two  random sequences
$t(\vec a,\lambda_a) = (1, 0, 0, 1, 1, 0, 0,\ldots$) and $t(\vec b,\lambda_b) =
(0, 0, 1, 1, 0, \ldots$).  The
nonlocality of
these correlations follows from the fact that whenever $t(\vec b, \lambda_b)
= 1$ on the
polarizer $ b$, we must have $t(\vec a, \lambda_a) = 1$ or 0 on the polarizer
$ a$ with the
probabilities  $P(1|1)$ and $P(0|1)$, i.e., the outcomes on $ a$ (possibly
even a
remote analyzer) are determined by the outcomes on the analyzer $ b$.
This is how the EPR quantum sequences of random numbers violate local
realism.
Only for $\alpha=0$ and $\alpha=\pi$ we have $P(0|0)=0$ and $P(1|1)=1$ and
the events do occur with certainty. In this case the EPR correlations are
perfect and the Bell's inequality is not violated.
In Figure~1 and Figure~2 are plots of the conditional
probabilities $P(1|0)$ and $P(0|1)$ as functions of the angle $\alpha$ for
different values of the spin $s$.  It is clear from these figures that
with the increased value of the spin $s$, the probability $P(1|0)$ tends to
zero while the probability  $P(0|1)$ tends to one except for the angle $\alpha
= \pi$, when $P(0|1)=0$. In this limit the random and  nonlocal character of
quantum mechanics goes away, and if $ s \to \infty$ we have:

\begin{equation}
P(\lambda_a|\lambda_b)  =  \left[ \matrix {1 &1 \cr 0 &0
\cr } \right]  \ \ \ {\rm for}\ \ \ \alpha   \neq \pi,
\end{equation}
and
\begin{equation}P(\lambda_a|\lambda_b) =  \left[ \matrix {1 &0 \cr 0 &1
\cr } \right] \ \ \ {\rm for}\ \ \ \alpha =\pi.
\end{equation}
This means that for $\alpha=\pi$ we have a 100\% confidence that the
outcome on $ a$ will be
the same as the outcome on $ b$ and  for  $\alpha  = 0$ we now have 0\%
confidence that the outcomes on $ a$ will be the same as the outcomes on $
b$.  This is precisely what we can expect
from the entangled correlations in the classical  limit corresponding in this
case  to
$ s \to \infty$. In this limit the quantum nonlocal distribution becomes a
local probability distribution corresponding
to sharp "yes" and "no" outcomes of the measurement involving classical
antiparallel angular momenta. In quantum mechanic terms, in this limit  the
quantum amplitudes become classical probabilities for a classical spining top
\cite{kw2,kw3}.

\subsection{RANDOMNESS AND INFORMATION}

  We have shown that the outcomes of the EPR correlations
can be visualized as nonlocal correlations between completely random
sequences
of random numbers $\lambda_a$ and $\lambda_b$.  The transitions between these
random numbers are described by joint and conditional probabilities given by
Eq.~(\ref{cp1}).  We have shown that due to the nonlocal character of these
transition probabilities the local reality of the spin correlations is
incompatible with the quantum mechanical predictions.  The nonlocality of the
transition probabilities (\ref{cp1}) violates the Bell's inequality.

On the other hand
random sequences $\lambda_a$ and $\lambda_b$ with their correlations provide
information about the correlated system of two spins. The joint or the
conditional probabilities for these random sequences
reflect what we know about the correlated system.  This knowledge called
information, can be described by the Gibbs-Shannon information-entropy H (in
bits). From the definition of joint and marginal distribution functions we
can
construct the following information entropies for the EPR correlations

\begin{equation}
H (a;  b) = - \sum_{\lambda_a\lambda_b} p(\lambda_a  \vec a; \lambda_b \vec
b)
\log_2 p (\lambda_a \vec a; \lambda_b \vec b),
\end{equation}
and its marginals:

\begin{equation}
H( \vec a) = - \sum_{\lambda_{a}} p(\vec a\lambda_a ) \log_2 p(\vec
a\lambda_a ) , \
H(\vec b) = - \sum_{\lambda_{b}} p(\vec b\lambda_b ) \log_2 p(\vec
b\lambda_b ).
\end{equation}
In the same way we introduce a conditional information entropy by the
following definition
\begin{equation}\label{hc}
H(\vec a| \vec b) = - \sum_{\lambda_a \lambda_b} p (\lambda_a \vec a;
\lambda_b
\vec b) \log_2 p (\lambda_a \vec a |\lambda_b \vec b) .
\end{equation}
{}From the Bayes's theorem (\ref{bayes1}) relating a conditional probability
with a joint
probability we obtain that
\begin{equation}
H ( \vec a;  \vec b )= H(\vec a | \vec b) + H ( \vec b)=  H(\vec b | \vec a)
+ H ( \vec a).
\end{equation}
{}From the properties of the joint and the conditional probabilities follow the
following inequality:
\begin{equation}\label{in1}
H(\vec a | \vec b) \leq H ( \vec a) \leq H ( \vec a;  \vec b) .
\end{equation}
This inequality has a simple interpretation that removing a condition never
decreases the information carried by the system.

For a system involving two correlated subsystems  the entropy
can offer a measure of quantum correlation if the information each  of the
subsystems is compared with the total information of the composite
system.  A very important property of the information-entropy is its
subadditivity.  The subadditivity states that \cite{wehrl}:
\begin{equation}\label{sub}
H ( \vec a;  \vec b) \leq H ( \vec a) + H ( \vec b)
\end{equation}
i.e, that when forming marginals one loses the information about the
correlations.
 Before discussing some applications of subadditivity, let us make some
remarks about the monotonicity of entropy.  Neither quantum-mechanically nor
in  classical physics is it true that $H( \vec a ) \leq H( \vec a;  \vec b)$.
This failure of monotonicity is expressed by the  Araki and Lieb (AL)
triangle inequality  \cite{alieb}:

\begin{equation}\label{al}
|H(\vec a) - H (\vec b) | \leq H (\vec a;\vec b) \leq
 H (\vec a) + H ( \vec b).
\end{equation}
with the right-hand side representing the subadditivity of the entropy.

{}From the EPR marginals (\ref{singlemarg})  we obtain:
\begin{equation}
H(\vec a) = H(\vec b) = -\frac{1}{2s+1} \log_2
\frac{(2s)^{2s}}{(2s+1)^{2s+1}} , \end{equation}
i.e., the information contained in the single spin measurements for the two
subsystems is equal to one bit for $s=\frac {1}{2}$ and is zero if
$ s \to \infty$. This follows from the fact that for $s=\frac {1}{2}$ we
have $P(0)=P(1)=\frac {1}{2}$ and for $ s \to \infty$ we have $P(0)=1$ with
certainty. With these marginal informations  the AL inequality (\ref{al})
takes the following form
\begin{equation}\label{al1}
0 \leq H (\vec a; \vec b) \leq -\frac{2}{2s+1} \log_2
\frac{(2s)^{2s}}{(2s+1)^{2s+1}} ,
\end{equation}
where the joint entropy for the information accumulated by the two polarizers
$\vec a$ and $\vec b$ is a function of  $\cos \alpha= \vec a \cdot  \vec b
$.  Let us note that this entropy is different from the von Neumann
definition involving ${\rm Tr} \{ \rho
\ln \rho \}$.  The EPR combined state is a pure state and as a result the von
Neumann entropy will be equal to zero.  The Gibbs-Shannon entropy as defined
by Eq.~(\ref{hc}) is different.  It involves statistical information associated
with possible ``states'' of the EPR system, labeled by indices $\lambda_a$
and
$\lambda_b$ and described by a transition probability $P(\lambda_a
|\lambda_b)$ to find the system in $\lambda_a$ if the system has been in
$\lambda_b$.

     In Figure~3 and Figure~4 we have plotted the conditional $H (\vec a|
\vec b)$, the joint   $H (\vec a; \vec b)$ and the single  $H (\vec a)$
folded information entropies as a function of the angle $\alpha$ for
spin-$\frac{1}{2}$ (Figure~3) and spin-$2$ (Figure~4).  We see that the AL
inequality (\ref{al1}) and the inequality (\ref{in1}) are satisfied for all
values of $\alpha$.  From
this plot we see that the additivity of entropy (i.e., $H(\vec a; \vec b) =
H(\vec a) + H(\vec b))$ holds only for $\alpha = \pi $, i.e., when $ \vec a$
is antiparallel to $ \vec b$.  This can be easily understood on the basis of
the transition probabilities.  For this geometry the only nonvanishing joint
probabilities are $P(1|1)=P(0|0)=1$ and as a result  $P(1;1)=P(1)$ and
$P(0;0)=P(0)$.  For the EPR correlations we conclude that the information
satisfies the
monotonicity condition, i.e., $H(\vec a) \leq H(\vec a;\vec  b)$.  Because of
this property the AL inequality and the curves from the Figure~3 and the
Figure~4 exhibit the subadditivity of the Gibbs-Shannon entropy when applied to
the nonlocal transition probabilities  $P(\lambda_a|\lambda_b)$.

Based on objective local realism, Braunstein and Caves (BC)
\cite{bcaves1,bcaves2} have derived the
information Bell's inequality involving the conditional  entropies for pairs
of probabilities. The BC inequality has the form:
\begin{equation}\label{bc1}
H(\vec a |\vec b ) \leq H(\vec a | \vec b' ) + H( \vec a ' | \vec b') +
H (\vec a' |\vec b ) ,
\end{equation}
and has the following simple physical interpretation, that four objective
quantities cannot carry less information than any two of them.
We shall apply this inequality to the information entropy that corresponds to
the joint and conditional
probabilities (\ref{cp1}) predicted by quantum mechanics for the EPR
correlations. For a coplanar geometry  $a\cdot \vec b'= \vec a'\cdot \vec
b'=\vec a'\cdot \vec b =\cos\alpha$ and $\cos3\alpha= \vec a\cdot \vec b$ the
BC inequality (\ref{bc1}) takes a
simpler form:

\begin{equation}
H(3\alpha ) - 3 H (\alpha)  \leq 0 ,
\end{equation}
where by $H(\alpha)= H(\vec a | \vec b ')$ we have denoted the conditional
entropy  as a
function of the angle $\alpha$.  In Figure~5 we have plotted the information
difference from the left hand
side of this inequality for different spins $s$.  From these curves we conclude
that the information Bell's inequality (\ref{bc1}) as presented by BC is
violated and that the
violations decrease with the spin value $s$. Note that the degree of
violation and the behavior of this violation for large values  of the spin is
different from the conclusions reached in \cite{bcaves1}. This is due to the
fact that in this reference the information entropy has been calculated for
measurements which resolve consecutive values of $m$, and such measurements
are inherently nonclassical in nature. The measurements presented in this
paper and represented buy the projections (\ref{projection}) do not resolve
values of $m$ and as a result do have a well defined classical limit.

\section{ENTANGLED GHZ CORRELATIONS }

\subsection{ RANDOMNESS AND NONLOCALITY}

 One particularly simple entangled GHZ
state of three spin-$\frac{1}{2}$ particles, named $a$, $b$, and $c$
discussed by Mermin \cite{mermin3} has the following form:
\begin{equation}\label{ghz}
|\psi_{\rm GHZ} \rangle =
\frac{1}{ \sqrt{2}} (|+, +, +\rangle  - | -, -, -\rangle) ,
\end{equation}
 where $|+\rangle$ or $|-\rangle $ specifies spin up or down along the
appropriate $z$-axis.  As showed by Mermin this entangled state provides an
{\it always} versus {\it never} test of local realism.
The joint probability for detection of the three particles $a$, $b$, and
$c$ by three  polarizers 1, 2 and 3, respective in an x-y plane perpendicular
to
the particles' line of flight is:
\begin{eqnarray}
 p(\phi) = p(\phi_1;\phi_2;\phi_3)=\langle \psi_{\rm GHZ}| \hat P(\phi_1)
\otimes \hat P(\phi_2) \otimes \hat P(\phi_3)|\psi_{\rm GHZ} \rangle
\nonumber\\
  = \frac {1}{ 8} (1 - \cos (\phi_1 + \phi_2 + \phi_3)) ,
\end{eqnarray}
 where $\phi_1, \phi_2$ and $\phi_3$ represent the orientation angles of the
detectors and $\phi= \phi_1+ \phi_2 + \phi_3$.  The case where a definite
prediction of spin measurement is
possible corresponds to $\phi_1 + \phi_2 + \phi_3 = \pi$ with the joint
probability equal to $p(\phi=\pi)= \frac{1}{4}$ and to $\phi_1 + \phi_2 +
\phi_3 = 0$ with the joint probability equal to $p(\phi=0)= 0$ .  These two
cases correspond to perfect correlations,
i.e., to such correlations when by measuring two spins one can predict with
certainty the outcome of the measurement involving the third spin. Contrary
to the EPR correlations, the GHZ correlations exhibit strong nonlocal
properties for perfect correlations.
       While for the two-particle EPR spin-singlet state, perfect
correlations can be made compatible with a stochastic model of
hidden-variable theory based on local realism, the three-particle GHZ perfect
correlations offer a {\it never} versus {\it always} refutation of local
realities.   It has been show that the GHZ correlations can be analyzed in the
context of cavity quantum electrodynamics,  and that the joint three-particle
probability can be measured, in principle, by   using single photon detection
\cite{kw4}.

Following the basic  ideas of a  theory based on local reality, the
transmission of an arbitrary spin-$s$ through  linear polarizers $\phi_1$,
$\phi_2$
and $\phi_3$ are described by objective realities with hidden variables
$\lambda_a$, $\lambda_b$ and $\lambda_c$.  In a theory based on LHV, as in
the EPR case (\ref{lhv}), the three-particle GHZ correlations are given  in
the form of the following statistical averages of local realities:
\begin{equation}\label{lhv1}
p(\phi_1;\phi_2;\phi_3) = \int d\lambda_a\int d \lambda_b \int d \lambda_c \
P(\lambda_a; \lambda_b;\lambda_c) \ t(\phi_1, \lambda_a)
\ t(\phi_2, \lambda_b) \ t(\phi_3, \lambda_c),
\end{equation}
where the hidden parameters are randomly distributed with a  positive and
normalized distribution function.
For such a local distribution of random variables  the  LHV transmission
probability (\ref{lhv1}) evaluated for four different polarization axes
$\phi_1,\ \phi_1',\ \phi_2,\ \phi_2'$ and a fixed $\phi_3$  is  restricted by
the following simple generalization of the  Bell's inequality ({\ref{bell1}})
\begin{eqnarray}\label{bell2}
-p(\phi_3) \leq  p(\phi_1; \phi_2;\phi_3) + p(\phi_1'; \phi_1;\phi_3) +
p(\phi_1; \phi_2';\phi_3)
- p(\phi_1'; \phi_2';\phi_3) \nonumber\\
 - p(\phi_1;\phi_3)- p(\phi_2;\phi_3) \leq 0 ,
\end{eqnarray}
Using the spectral decomposition for the three GHZ projector operators
 the three-particle GHZ correlations can be written in the following
form\begin{equation}
 p(\phi) =
\int d \lambda_a \int d \lambda_b \int d \ \lambda_c \  \ P(\phi_1
\lambda_a;\phi_2 \lambda_b;\phi_3 \lambda_c)\ \lambda_a \ \lambda_b
\ \lambda_c  ,
\end{equation}
where the distribution function is defined as
\begin{equation}
 P(\phi_1 \lambda_a;
\phi_2\lambda_b; \phi_3\lambda_c) = \langle\psi_{\rm GHZ} | \delta (\lambda_a
- \hat P(\phi_1)) \otimes \delta
(\lambda_b - \hat P(\phi_2))  \otimes \delta (\lambda_c - \hat
P(\phi_3))|\psi_{\rm GHZ} \rangle .
\end{equation}
The relation of the randomness and the nonlocality of the GHZ correlations
will be  addressed with a statistical analysis of the three-particle
distribution
function. We discuss these  correlations in the framework of Bayes analysis
using  the three-point generalization of the the two-point EPR
(\ref{nonlocp1}) distribution \cite{kw5}.
For this distribution we can write:
\begin{equation}\label{bayes2}
P(\phi_1 \lambda_a;
\phi_2\lambda_b; \phi_3\lambda_c) = P (\phi_1\lambda_a |
\phi_2\lambda_b \phi_3\lambda_c ) P (\phi_2\lambda_b |\phi_3\lambda_c) P
(\phi_3\lambda_c) ,
\end{equation}
where the distribution $P (\phi_2\lambda_b |\phi_3\lambda_c)$ is the
conditional of the
event $\lambda_b$ ({\it yes} or {\it no}) to occur under the condition that
$\lambda_c$ ({\it yes} or {\it no}) has occurred.  The distribution
$P(\phi_1\lambda_a |\phi_2\lambda_b \phi_3\lambda_c)$ is the conditional of
the event $\lambda_a$
to occur under the condition that $\lambda_b$ and $\lambda_c$ have occurred.
The distribution $P(\phi_3\lambda_c)$ is a one-fold marginal of
Eq.~(\ref{bayes2}).  The three-particle distribution function is positive
everywhere, but depends on the polarization
directions $\phi_i$. The nonlocality of the GHZ  distribution
function makes the Bell's inequality (\ref{bell2}) incompatible with quantum
prediction. As in the previous section, for notational convenience, we will
omit the polarization directions and denote these functions by $P(i|jk)$,
$P(i|j)$ and $P(j)$, where
the indexes $i$ $j$ and $k$ will denote $1$ and $0$ for a "yes" or "no"
outcomes.  From the
GHZ spin state (\ref{ghz}) we obtain that:  $P(\lambda_c) = \frac {1}{2}$ and
$P (\lambda_b
|\lambda_c) = \frac{1}{2}$ for all values of $\lambda_b$ and $\lambda_c$ and
all orientations of the polarizers.  The  the three-fold conditional
probabilities are

\begin{eqnarray}\label{pc2}
P(1|11) = P(0|01) = P(1|00) = P(0|10) = \frac{1}{2}(1-\cos\phi), \nonumber\\
P(0|00) = P(1|10) = P(1|01) = P(0|11) = \frac{1}{2}(1+\cos\phi)
\end{eqnarray}
These last expressions reveal the statistical nature of GHZ correlations,
that the value of the third spin is known with probability
$\frac{1}{2}(1\pm\cos\phi)$  if the other two
spins have be measured.  This statistical picture leads to a simple
interpretation of GHZ correlations in terms of random numbers 1 and 0.  The
quantum mechanical average in this case is represented by an ensemble average
of three sequences of random numbers 1 and 0, that are the only possible
outcomes of the experiment.  On each single polarizer or  pair of polarizers
the outcomes are completely random and the {\it yes} and {\it no}  answers
occur with equal probability $P(\lambda_a |\lambda_b)=\frac{1}{2}$.  The  GHZ
correlations show that these completely random sequences are correlated if
a third  polarizer is involved. For example for $\phi =0$ if the two
detectors have registered $(0,0)$ the third detector has to register  the
outcome $0$ with certainty. For $\phi =\pi$,  if the two detectors have
registered $(1,1)$ the third detector has to register  the outcome $1$ with
certainty. Any subsequence of random numbers composed only of pairs
$(\lambda_a ,\lambda_b)$, $(\lambda_a ,\lambda_c)$ and $(\lambda_b ,\lambda_c)$
is completely random and uncorrelated, while sequences involving three outcomes
$(\lambda_a, \lambda_b, \lambda_c)$ are correlated and are described by a
nonlocal-polarization-dependent
distribution function.  Note that the
three-fold probability distribution (\ref{pc2}) fails to satisfy the property
that the probability of an event like the registration of each measurement
depends only on the preceding outcome and not also on earlier states of the
measured system.  In short we have:

\begin{equation}
P (\lambda_a|\lambda_b \lambda_c) \not= P
(\lambda_a|\lambda_b)P (\lambda_b |\lambda_c) .
\end{equation}
These conclusions are valid for any an arbitrary orientation of the
polarizers and indicates that strong nonlocality is exhibited even for
perfect correlations corresponding to $\cos\phi = \pm 1$.

\subsection{RANDOMNESS AND INFORMATION}
 The outcomes of the GHZ
correlations can be visualized as nonlocal correlations between completely
random sequences  of random numbers $\lambda_a$ and $\lambda_b$ and
$\lambda_c$.  The transitions between these
random numbers are described by joint and conditional probabilities given by
Eq.~(\ref{pc2}).  Due to the nonlocal character of these
conditional probabilities the local reality of the GHZ  spin correlations is
incompatible with the quantum mechanical predictions. The nonlocality of
the transition probabilities violates the Bell's inequality.  On the other
hand random sequences $\lambda_a$ $\lambda_b$ and $\lambda_c$ with their
correlations provide information about the correlated system of the three
spins. The joint or the conditional probabilities for these random sequences
reflect what we know about the GHZ correlated system.  Following the
definitions of the information entropy for the EPR correlations  we
introduce  the following information entropy for the GHZ correlations:

\begin{equation}
H (\phi_1 ;\phi_2 ;\phi_3 ) = - \sum_{\lambda_a\lambda_b\lambda_c} P(\phi_1
\lambda_a;\phi_2 \lambda_b;\phi_3 \lambda_c) \log_2  P(\phi_1
\lambda_a;\phi_2 \lambda_b;\phi_3 \lambda_c).
\end{equation}
For a system involving three correlated subsystems   the entropy
can offer a measure of quantum correlation if the informations of the
subsystems $\phi_1 \cup \phi_3$ and $\phi_2\cup \phi_3$ are compared with the
informations of the subsystems $\phi_1$ and $\phi_3$. For such  measurements
Lieb and Ruskai (LR)\cite{liebrus} have derived the following inequality:
\begin{equation}
H (\phi_1 ;\phi_3 ) - H (\phi_1 ) + H (\phi_2 ;\phi_3 )-\ H (\phi_2 ) \geq 0.
\end{equation}
The LR inequality implies that more information can be obtained in joint
measurements involving various two-particle correlations than in single
measurement involving only one particle. For the GHZ correlations, $H (\phi_1
;\phi_3 )=H (\phi_2 ;\phi_3 )=2$ and $ H (\phi_1 )= H (\phi_2 )=1$ and we see
that this inequality is satisfied independently of the polarization
orientations.
 For the three-particle system the subadditivity for the entropy difference
takes the following form:

\begin{equation}
H (\phi_1 ;\phi_2 ;\phi_3 ) - H (\phi_2 ) \leq  H (\phi_1 ;\phi_2 ) - H
(\phi_2) + H (\phi_3 ;\phi_2 )-H (\phi_2 ).
\end{equation}
This inequality implies that the information content of $\phi_1 \cup \phi_2$
and $\phi_3 \cup \phi_2$ relative to $\phi_2$ is greater than that of $\phi_1
\cup \phi_2 \cup \phi_3$ relative to $\phi_2$.
For the GHZ wave function this inequality reduces to:
\begin{equation}
H(\phi)= H (\phi_1 ;\phi_2 ;\phi_3 )\leq 3.
\end{equation}
As in the case of the LR inequality, this inequality is satisfied for all
values of the the angle $\phi$. In Figure~6 we see a plot of the joint
three-particle information entropy as a function of the polarizers angles
forming $\phi$. We see that the the maximum information is reahced if $\phi_1
+\phi_2 +\phi_3 = \pi $.

Based on  objective local
realism we generalize the BC information Bell's
inequality (\ref{bc1}) involving information entropies of the GHZ  three
particles. The three-particle BC information Bell's
inequality for the GHZ correlations has the form

\begin{equation}\label{bc2}
H(\phi_1 |\phi_2 \phi_3) \leq H(\phi_1  |\phi_2' \phi_3) + H( \phi_1 '
|\phi_2' \phi_3) +
H (\phi_1' |\phi_2 \phi_3) ,
\end{equation}
where $H(i|jk)$ are the conditional three-point distributions $P(i|jk)$
given by Eqs.(\ref{pc2}). In order to investigate the properties of this
inequality, we select the following orientations of the polarization angles;
$\phi'_1 = \frac{3\pi}{4}$, $\phi'_2 = \frac{\pi}{2}$, $\phi_1 = \frac{\pi}{4}$
and $\phi_2 = 0$ (see Figure~7). For this particular geometry the
information Bell's inequality (\ref{bc2}) for the GHZ correlations reduces to:
\begin{equation}\label{bc3}
H(\frac{\pi}{4}+\phi_3 ) - 3 H(\frac{3\pi}{4}+\phi_3 ) \leq,
\end{equation}
In Figure~8 we have plotted the information difference represented by the left
hand side of this inequality.  From
this curve we conclude that the information Bell's inequality (\ref{bc3}) as
presented by BC is violated.

\section*{Acknowledgments}

The author thanks C. Caves and G. Herling for
numerous discussions and comments.
This work was partially supported the Center for Advanced Studies of the
University of New Mexico.

\newpage

\begin{figure}
\caption[f1]{
The conditional distribution function $P(1|0)$ as a function of the angle
$\alpha$ for different vaules of the spin. Spin-$\frac{1}{2}$ curve (a),
spin-$1$ curve (b), spin-$2$ curve (c), spin-$5$ curve (d).}
\end{figure}

\begin{figure}
\caption[f2]{
The conditional distribution function $P(0|1)$ as a function of the angle
$\alpha$ for different vaules of the spin. Spin-$\frac{1}{2}$ curve (a),
spin-$1$ curve (b), spin-$2$ curve (c), spin-$5$ curve (d).}
\end{figure}

\begin{figure}
\caption[f3]{
Plots of the  information entropies $H(\vec a | \vec b )$ curve (a),  $H(\vec a
)$ curve (b) and   $H(\vec a ; \vec b)$ curve  (c) as functions of the angle
$\alpha$  for spin $s=\frac{1}{2}$.}
\end{figure}

\begin{figure}
\caption[f4]{  Plots of the information entropies $H(\vec a | \vec b )$ curve
(a),  $H(\vec a )$ curve (b) and  $H(\vec a ; \vec b)$ curve  (c)  as functions
of the angle $\alpha $  for spin $s=2$.}
\end{figure}

\begin{figure}
\caption[f5]{
Plots of the information difference in the BC Bell's inequality as functions of
the angle $\alpha$ for for different values of the spin. Spin-$\frac{1}{2}$
curve (a), spin-$1$ curve (b), spin-$2$ curve (c), spin-$5$ curve (d).}
\end{figure}

\begin{figure}
\caption[f6]{
Plot of the three-particle GHZ information as a function of the polarizers
angle $\phi$.}
\end{figure}

\begin{figure}
\caption[f7]{
The coplanar orientations of the three polarizers with $\phi'_1 =
\frac{3\pi}{4}$, $\phi'_2 = \frac{\pi}{2}$, $\phi_1 = \frac{\pi}{4}$ and
$\phi_2 = 0$.}
\end{figure}

\begin{figure}
\caption[f8]{
Plot of the three-particle GHZ  information difference following from the
Bell's inequality as functions of the angle $\phi_3$.}
\end{figure}




\end{document}